\documentclass[usenatbib]{mnras}
\usepackage{amsmath}
\usepackage{amssymb}
\usepackage{graphicx}
\usepackage{grffile}
\usepackage{float}
\usepackage[dvips]{epsfig}
\usepackage{epsfig}
\usepackage{dblfloatfix}
\usepackage{color}
\usepackage{caption}
\usepackage{bm}
\usepackage[british]{babel}

\def\({\left(}
\def\){\right)}
\def\[{\left[}
\def\]{\right]}

\begin{document}

\title[$\beta$-Skeleton Analysis]{$\beta$-Skeleton Analysis of the Cosmic Web}
\author[Fang, Forero-Romero, Rossi, Li \& Feng (2018)]
{Feng Fang$^1$, Jaime Forero-Romero$^2$, Graziano Rossi$^3$, Xiao-Dong Li$^{1,\star}$, Long-Long Feng$^1$ \\ \\
$^1$ School of Physics and Astronomy, Sun Yat-Sen University, Guangzhou 510297, P. R. China \\
$^2$ Departamento de F{\'i}sica, Universidad de los Andes, Cra. 1 No. 18A-10 Edificio Ip, CP 111711, Bogot{\'a}, Colombia \\
$^3$ Department of Physics and Astronomy, Sejong University, Seoul, 143-747, Korea
$^\star$corresponding author: lixiaod25@mail.sysu.edu.cn}
\pagerange{\pageref{firstpage}--\pageref{lastpage}} \pubyear{2018}
\maketitle
\label{firstpage}


\begin{abstract}
The $\beta$-skeleton is a mathematical method to construct graphs from
a set of points that  has been widely applied in the areas of image
analysis, machine learning, visual perception, and pattern recognition.  
In this work, we apply the $\beta$-skeleton to study the cosmic web. 
We use this tool on observed and simulated data to identify the
filamentary structures and characterize the statistical properties of
the skeleton.  
In particular, we compare the $\beta$-skeletons built from SDSS-III galaxies to 
those obtained from MD-PATCHY mocks, and also to mocks directly built from the Big MultiDark $N$-body
simulation.
We find that the $\beta$-skeleton is able to reveal the underlying structures 
in observed and simulated samples without any parameter fine-tuning. 
 A different degree of sparseness can be obtained by adjusting the
value of $\beta$; in addition,  
the statistical properties of the length and direction of the
skeleton connections show a clear dependence on redshift space
distortions (RSDs) and galaxy bias.
We also find that the $N$-body simulation accurately reproduces
the RSD effect in the data, while the  MD-PATCHY mocks appear to underestimate
its magnitude. 
Our proof-of-concept study shows that the statistical properties of the
$\beta$-skeleton can be used to probe
cosmological parameters and galaxy evolution.
\end{abstract}



\section{Introduction}


The spatial distribution of the nearest galaxies on scales of a few
hundred Megaparsecs follows a distinct filamentary motif.
This pattern is known as `cosmic web'
\citep{1986ApJ...304...15B}, and it has been observed at
different cosmic epochs \citep{1986ApJ...302L...1D,
  2012ApJS..199...26H,2004ApJ...606..702T,2014A&A...566A.108G}.   
The search of consistent and stable methods to define this web-like structure has
been the subject of continuous research for the last $\sim 40$ years, since its
existence was confirmed in early cosmic maps from galaxy redshift
surveys.
The cosmic web has also been detected in  the dark matter description
provided by cosmological simulations --  
see \cite{2018MNRAS.473.1195L} for a recent review.

The cosmic web is usually classified into four different components: halos,
sheets, filaments, and voids.
Many algorithms are focused in finding the two most prominent web features
present in redshift galaxy surveys: voids and filaments. 
Voids are regions with sizes in the range of $20-50$ Mpc,
practically devoid of galaxies -- see \cite{2016IAUS..308..493V} for a recent review of
void finding algorithms.
Filaments appear to be the main bridges connecting high-density regions. 
On the largest scales, the filament length can be on the order of
$10-100$ Mpc.

The emergence of the cosmic web can be understood as the interplay of
two conditions. 
First, the initial Gaussian random density field; second, its
evolution under gravity.  
In fact, the initial anisotropies in the density field are amplified by gravity
to finally become filaments and voids.  
The structure of the cosmic web is thus expected to encode information
about the underlying cosmological model: namely, type of initial fluctuations,
proportions of different kinds of matter, the expansion history of the
Universe, and the rules of gravity.
Voids, for instance, can be used as cosmological probes, as their
structure is strongly influenced by dark energy \citep{2009ApJ...696L..10L,2012MNRAS.426..440B};
and the statistical isotropy of filaments can be used to perform the
Alcock-Paczynski (AP) test \citep{2014ApJ...796..137L}.

In this paper, we introduce the $\beta$-skeleton as an algorithm to
characterize the cosmic web.
The $\beta$-skeleton concept stems from the fields of computational geometry
and geometric graph theory and has been widely applied in the areas of
image analysis, machine learning, visual perception, and pattern
recognition \citep{edelsbrunner1983shape,amenta1998crust,zhang2002locating}. 
In the context of web finders, the $\beta$-skeleton belongs to a class
of algorithms that, starting from a set of 3D spatial points, builds a
graph describing the degree of connectedness.
In this aspect, it is similar to the minimum spanning tree (MST)
algorithm \citep{1985MNRAS.216...17B}, with the main difference that the resulting graph depends
on the continous $\beta$ parameter;
it is also related to web finders that are designed on the basis of
topological persistence, such as DisPerSE \citep{2011MNRAS.414..350S}. 

This paper is organized as follows. 
In Section 2, we briefly introduce the definition and the basic properties of the $\beta$-skeleton. 
In Section 3, we describe the Big MultiDark Planck (BigMDPL) simulation and the SDSS-III BOSS Data Release 12 (DR12) galaxy sample, which are used later on in the analysis. 
The application of the $\beta$-skeleton statistics is presented in Section 4, 
where we discuss the dependence of the skeleton on the values of $\beta$, 
on the redshift of the various samples, on the redshift-space distortions (RSDs), and on the cosmological volume and AP effects; 
we also graphically illustrate the $\beta$-skeleton constructed from SDSS-III BOSS DR12 galaxies,
and eventually compare the skeletons obtained from observational data and simulated catalogs.  
Finally, we summarize our findings and conclude in Section 5. 

 
\section{$\beta$-Skeleton: Theory}


In what follows, we 
define the $\beta$-skeleton and briefly explain how it is used to study the statistical properties of the large-scale structure (LSS) of the universe;   
for more details about the $\beta$-skeleton in topology and in geometric graph theory, 
please refer to \cite{kirkpatrick1985framework,correa2012locally}.

\begin{figure}
 \centering
 \includegraphics[width=7cm]{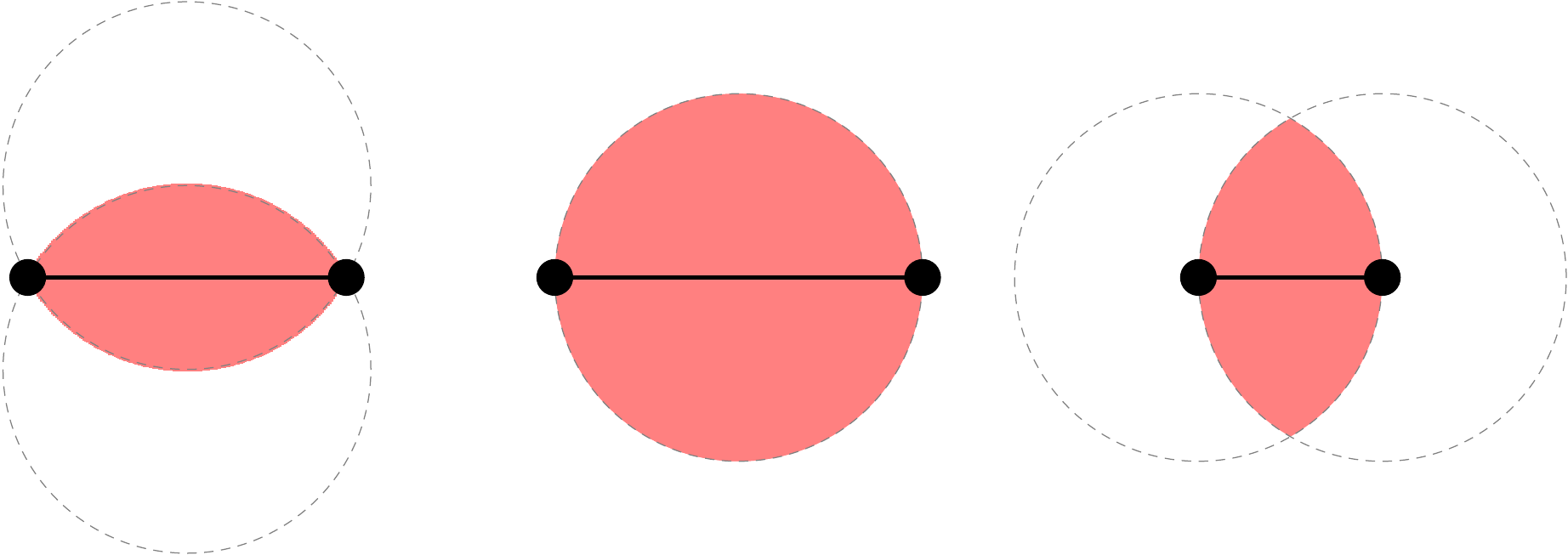}
 \caption{\label{bsk_define} Empty region of the $\beta$-skeleton under the Lune-based definition. Left: $\beta <$1, Middle: $\beta$=1, Right: $\beta > $1}.
\end{figure}

For a point set $S$ in a $n$-dimensional Euclidean space, 
the $\beta$-skeleton defines an {\it edge set} so that for any two points $p$ and $q$ in $S$, 
those points are considered to be
connected if there is not a third point {\it r} in the various $empty\ regions$ shown in Figure \ref{bsk_define} with dotted lines.
Specifically: 

\begin{itemize}
 \item For $0<\beta<1$, the {\it empty region} is the intersection of all the spheres with diameter ${d_{\rm pq}}/{\beta}$, having $p$ and $q$ on their boundary. 
 \item For $\beta = 1$, the {\it empty region} is the sphere with diameter $d_{\rm pq}$.
 \item For $\beta \geq 1$, the {\it empty region} is defined in two different ways:  namely, the {\it Circle-based} definition and the {\it Lune-based} definition (see again Figure \ref{bsk_define} for details). 
In this paper, we adopt the latter one, according to which the {\it empty region}
$R_{\rm pq}$ is the intersections of two spheres with diameter $\beta d_{\rm pq}$ and centered at $p+\beta(q-p)/2$ and $q+\beta(p-q)/2$, respectively.
\end{itemize}

The $\beta$-skeleton defined above has several interesting mathematical properties. 
As $\beta$ varies continuously from 0 to $\infty$,
the constructed graphs change from a complete graph to an empty graph. 
The special case of $\beta$ = 1 leads to the so called  "Gabriel graph",
which is known to contain the Euclidean minimum spanning tree.\footnote{
 minimum spanning tree of a set of $n$ points in the plane
where the weight of the edge between each pair of points
is the Euclidean distance between those two points.} 
The $\beta$-skeleton has several 
important applications in computational science and graphical theory. 
For example, in image analysis, it was used to reconstruct the shape of a two-dimensional object given a set of sample points on the boundary of the object: 
this is because the $\beta=1.7$ {\it Circle-based} graphs have been proven to correctly reconstruct the entire boundary of any smooth surface, without generating  
any edges that do not belong to the boundary --
as long as the samples are sufficiently dense with respect to the local curvature of the surface.\footnote{In experimental testing, $\beta= 1.2$ 
was more effective in reconstructing street maps from a set of points,  
marking the center lines of the streets in a geographic information system.} 
The $\beta$-skeleton has also been applied in machine learning systems, in order  
to solve geometric classification problems \citep{zhang2002locating,toussaint2005geometric}. 
In wireless ad hoc networks, for controlling the communication complexity,
the $\beta$-skeleton was used as a mechanism  
to choose a subset of the pairs of wireless stations
that can communicate with each other \citep{bhardwaj2005distributed}.
In visual perception and pattern recognition,
it was used to find families of proximity graphs \citep{ersoy2011skeleton}.
For more details about the application of the $\beta$-skeleton, see e.g.
\cite{bose2002spanning,wang2008topology,lafarge2013surface}.

\section{Observed and Simulated Datasets}

First, we test our method using the BigMDPL simulation. 
The BigMDPL belongs to the series of MultiDark $N$-body simulations with Planck 2015 cosmology, thoroughly
described in \cite{klypin2016multidark}.
It is characterized by a box with $2.5 h^{-1}$Gpc on a side,  
with $3840^3$ dark matter particles, providing a mass resolution of $2.4 \times 10^{10}$ $h^{-1} M_{\odot}$.
The initial conditions, based on primordial Gaussian fluctuations, 
are generated via the Zel'dovich approximation at $z_{\rm init} = 100$. 
The cosmology assumed is a flat $\Lambda$CDM model with   
$\Omega_m = 0.307115$, $\Omega_b = 0.048206$, $\sigma_8 = 0.8288$, $n_s = 0.9611$, and $H_0 = 67.77~{\rm km}\ s^{-1} {\rm Mpc}^{-1}$.

We then apply the $\beta$-skeleton statistics to the Baryon Oscillation Spectroscopic Survey (BOSS) DR12 CMASS galaxy sample.
BOSS \citep{dawson2012baryon,smee2013multi}, is the cosmological counterpart of the Sloan Digital Sky Survey III (SDSS-III;  Eisenstein et al. 2011),
and it is still one of the largest spectroscopic galaxy surveys to date. It has obtained spectra and redshifts 
of about $1.37$ million galaxies selected from the SDSS imaging up to $z=0.7$.
The Northern and Southern Sky footprints cover an area of $\sim 10,000$ square degrees,
and the galaxy samples are conventionally split into the LOWZ  catalog at $z\leq0.43$ and 
the CMASS catalog covering the redshift interval $0.43\leq z \leq 0.7$ \citep{reid2015sdss}.
In this work, we only use the CMASS sample at $0.43\leq z \leq 0.7$, which contains $\sim 0.77$ million galaxies.

In order to compare observational data with $N$-body simulation predictions, we
use the MD-PATCHY mocks available for the BOSS survey.
The MD-PATCHY mocks \citep{kitaura2016clustering, rodriguez2016clustering} adopt an 
halo abundance matching technique to reproduce 
 the two- and three-point clustering measurements of BOSS. 
The redshift evolution of the biased tracers is matched to the corresponding observations by applying the 
aforementioned technique in a number of redshift bins,
with the resulting mock catalogs being combined together to form a contiguous lightcone.
The MD-PATCHY mocks are constructed to reproduce the number density, 
selection function, and survey geometry of the BOSS DR12 catalog; moreover, 
the two-point correlation function (2PCF) of the observational data is correctly recovered down to a few Mpc scales, 
in general within $1\sigma$ error \citep{kitaura2016clustering}.
The MD-PATCHY mocks have been carefully tested and subsequently adopted for the statistical analysis of BOSS data
in a series of works -- see for example \cite{alam2017clustering}, and references therein.


\section{Main Results}


\subsection{An Illustrative Application}

As an illustrative example, we first apply the $\beta$-skeleton statistics to a set of LSS mock samples using $\beta = 1,3,10$, respectively. 
We do this as follows: essentially, we simply take the $z = 0$ halo catalog of the BigMDPL simulation and apply a mass cut $M>1\times 10^{13} M_{\odot}$ 
and a radial cut $r_{cut} < 500 h^{-1} {\rm Mpc} $. This procedure allows us to 
create a shell-shaped sample containing $30,000$ dark matter halos. 
In order to make comparisons with an unclustered distribution, we also 
built a random sample with the same size, shape, and number of points as the previous mock realizations.  

\begin{figure*}
 \centering
 \includegraphics[width=15cm]{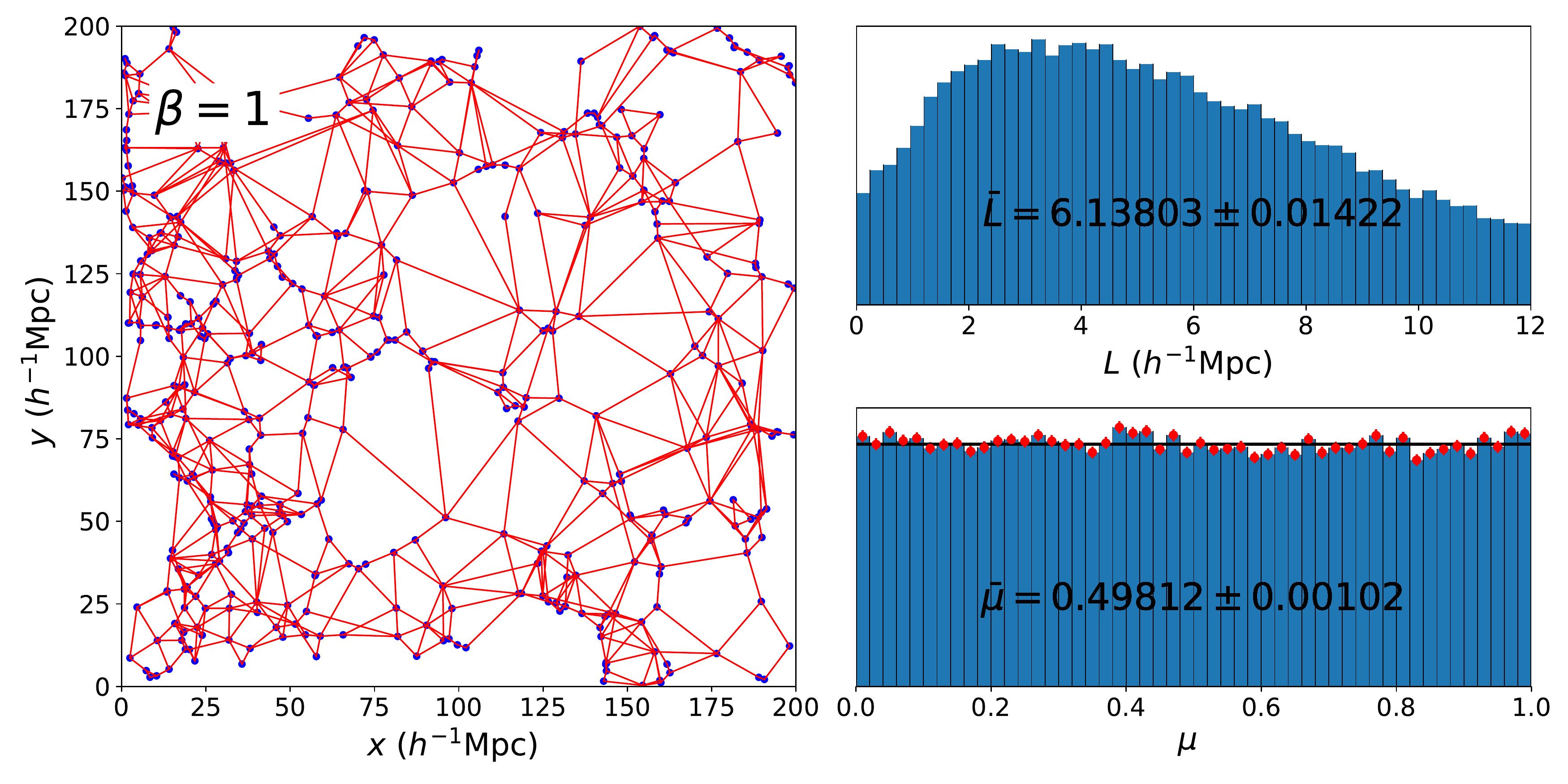}
 \includegraphics[width=15cm]{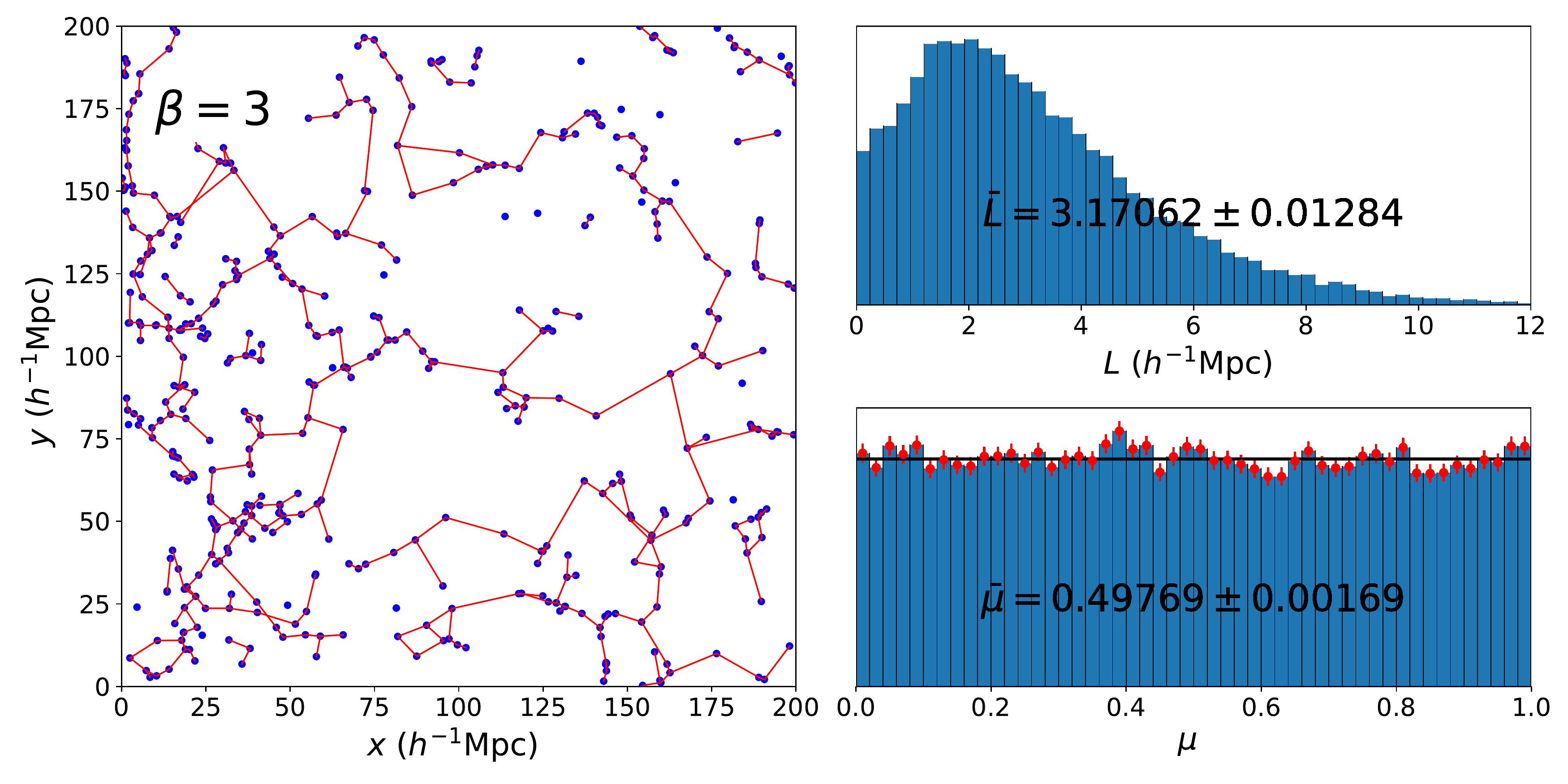}
 \includegraphics[width=15cm]{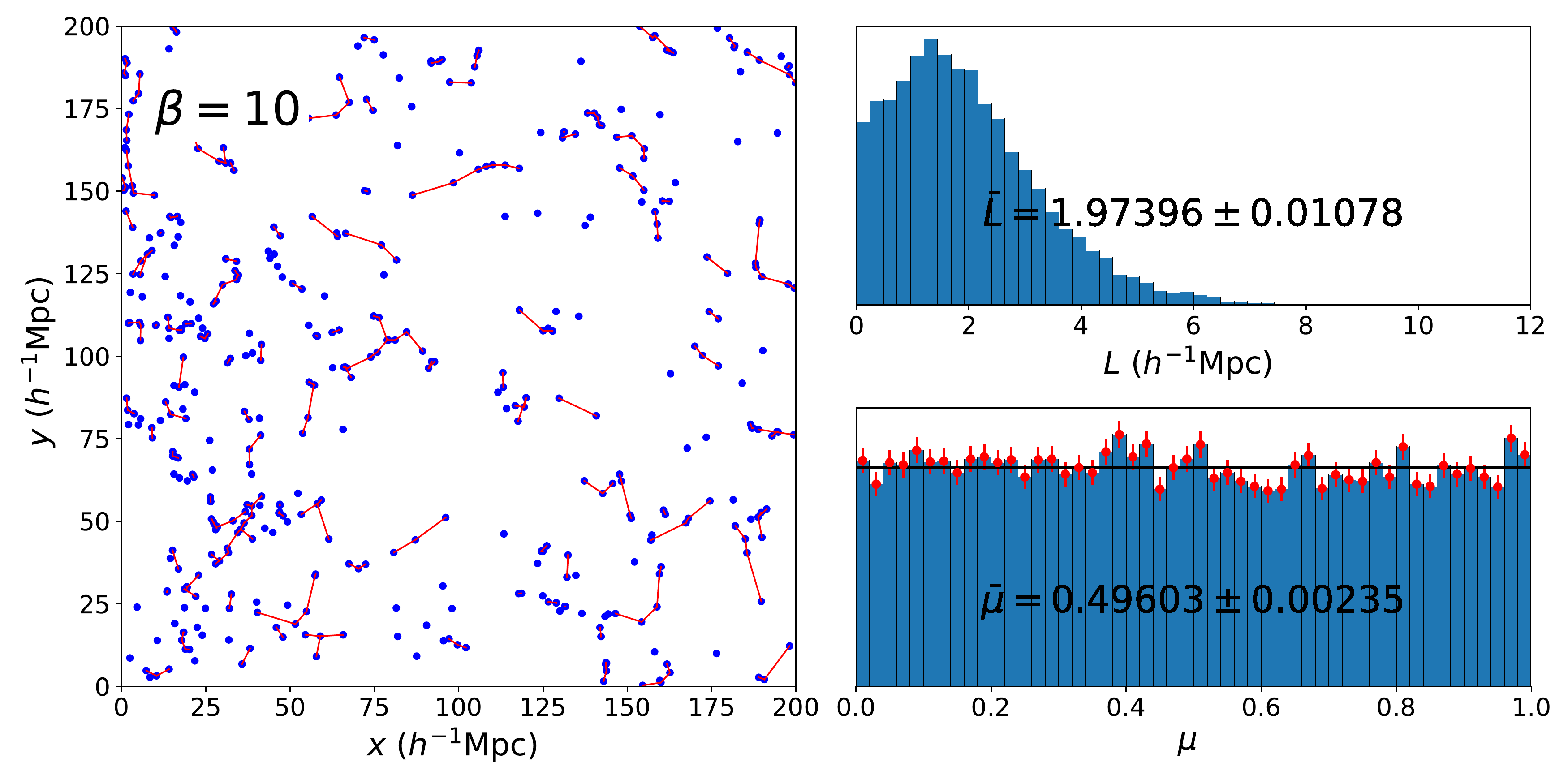}
 \caption{\label{beta_Nbody} {\it An illustrative example}. Application of the $\beta$-skeleton statistics to a set of LSS mock samples when $\beta = 1,3,10$, respectively. 
 In the figure, the left panels show the skeletons of the mock samples for different values of $\beta$, while the 
 right panels
present the statistics of the length of the connections (upper parts)
and the orientations of  those connections  (lower parts). See the main text for more details.  
 }
\end{figure*}

\begin{figure*}
 \centering
 \includegraphics[width=14cm]{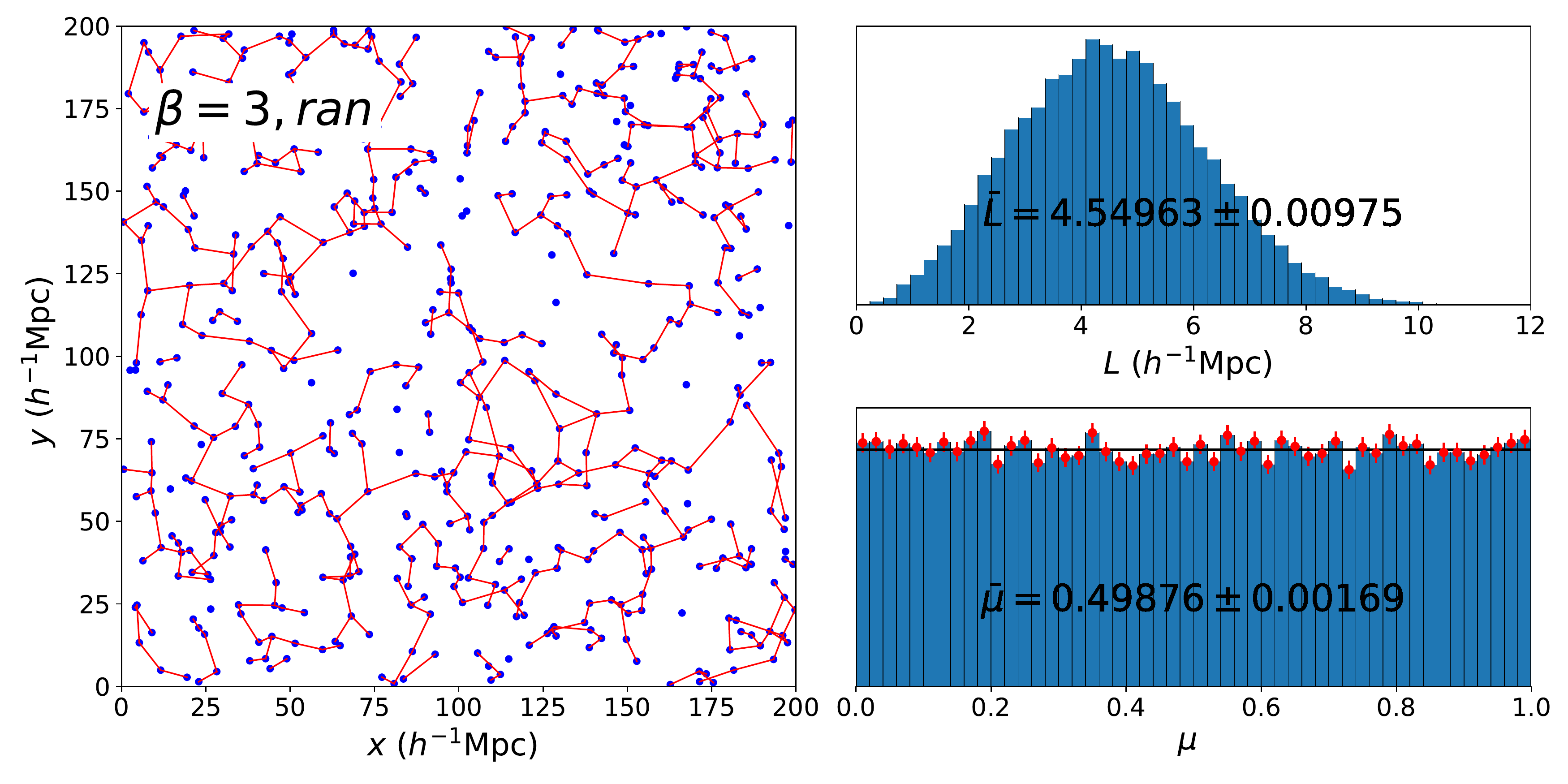}
 \caption{\label{beta_random} Same as the previous figure when $\beta=3$, but for an unclustered (random) distribution.}
\end{figure*}

Results of this test are displayed in Figures \ref{beta_Nbody} and \ref{beta_random}, 
where we show the skeletons of the mock samples using $\beta=1,3,10$, respectively (from top to bottom) -- as well as 
the skeleton of the random sample when $\beta=3$. 
In all cases, the left panels display a $200\times200\times30 ~h^{-1} {\rm Mpc}$ slice of the samples with connections (red lines), 
while the right panels
show histograms of the length of the connections $L$ (upper part)
and the cosine of the angle between the line-of-sight (LOS) and the connection line, $\mu\equiv |cos\theta|$ (lower part). 

Clearly, the amount of connections is smaller when $\beta$ is larger. 
This is evident from the definition for the $\beta$-skeleton presented in Figure \ref{bsk_define}, 
which shows an increment of the empty region with $\beta$; namely,  
the threshold for having two particles connected becomes more strict. 
In particular, when $\beta$ = 1 we find $\sim 80, 000$ connections, 
far more than the number of points of the sample, 
while we detect only $15,000$ connections when $\beta = 10$. 

The $\beta$-skeleton automatically generates filament-like structures from the point sample; 
this is most clearly detected when $\beta=3$, as can be seen in Figure \ref{beta_Nbody}. 
For example, in the upper-left panel one can notice that $\sim 20$ galaxies naturally arise from 
a long straight filament-shape structure: this structure is then identified, and those galaxies are linked together.
The straight line ends at $(x,y)\approx (25,140)~h^{-1}$ Mpc,
while the structure continues and extends up to $y=60~h^{-1}$ Mpc.
It then bifurcates at $y \approx 110~h^{-1}$ Mpc, and further extends to the left, lower-left, and right side of the graph, 
forming a larger connected structure which captures $\approx 70\%$ of the galaxies shown in the panel.
Also, some of those galaxies act as ``knots'' of the structure (i.e., three or more galaxies are connected). 
For example, the ``knots'' galaxy at $(x,y)\approx (25,140) h^{-1}$ Mpc links together the up-down filament at its left to the galaxies at the right.
Moreover, there are also isolated structures having a relatively 
small number of group members -- see for instance 
$\sim 25$ galaxies distributed around 
$(x,y)\approx (100,150) h^{-1}$ Mpc that form an ``A''-shaped structure.

Altering the values of $\beta$ has a strong influence on the 
overall shape of the skeleton graphs. For example, 
the case of $\beta=1$ roughly corresponds to computing
the 2PCF, in the sense that many connections are generated, regardless of whether or not those 
connections lie within a filament When  $\beta=3$, the set of structures generated is much closer to the observed cosmic web,
meaning that the number of connections is comparable to the number of actual galaxies.
For $\beta=10$, one gets a very sparse graph as expected, since 
only the small and relatively isolated compact groups of galaxies 
are identified and connected.

The statistical properties of the connection length $L$ also vary with $\beta$. 
For larger values of $\beta$, $L$  gets smaller and appears to be more concentrated  -- this is because, due to a tight threshold, it is difficult to connect two points 
separated by a large distance. From the figure, we infer that the mean length is $\bar L = 6.14, 3.17, 1.97 ~ h^{-1}$ Mpc when $\beta = 1,3,10$, respectively. 

For $\beta=3$, we then compare the results obtained from the mock samples with those derived 
from the unclustered (random) distribution. As expected, we find that
the random sample exhibits  ``structures'' chaotic in shape;  moreover, due to a lack of compact structures, the 
distribution of $L$ inferred from the random sample  has a mean $\bar
L=4.55$, a value much larger than those obtained from the mock
samples.  

Finally, as shown in all the bottom right panels of Figures  \ref{beta_Nbody}  and \ref{beta_random}, 
we find that  $\mu \sim 0.5$ within the corresponding errorbars, 
implying that the directions of the connections  are always randomly distributed,  with no preferred orientation. 
 
\subsection{Redshift Evolution}

Next, we study in detail the statistical properties of the $\beta$-skeletons constructed from $N$-body simulations. 
We analyze 4 BigMDPL snapshots at redshifts $0, 0.3, 0.6$, and $0.9$, respectively, and consider both cases with and without 
RSD effects. 
Moreover,  we impose a mass cut $M<1\times 10^{13} M_{\odot}$
and a radial cut $r<2500 ~h^{-1} {\rm Mpc}$,
yielding a number of galaxies $N_{\rm gal}=$ 
3.85, 3.37, 2.71, 2.07 (in units of millions) at those 4 redshifts, respectively.

\begin{figure*}
 \centering
 \includegraphics[width=18cm]{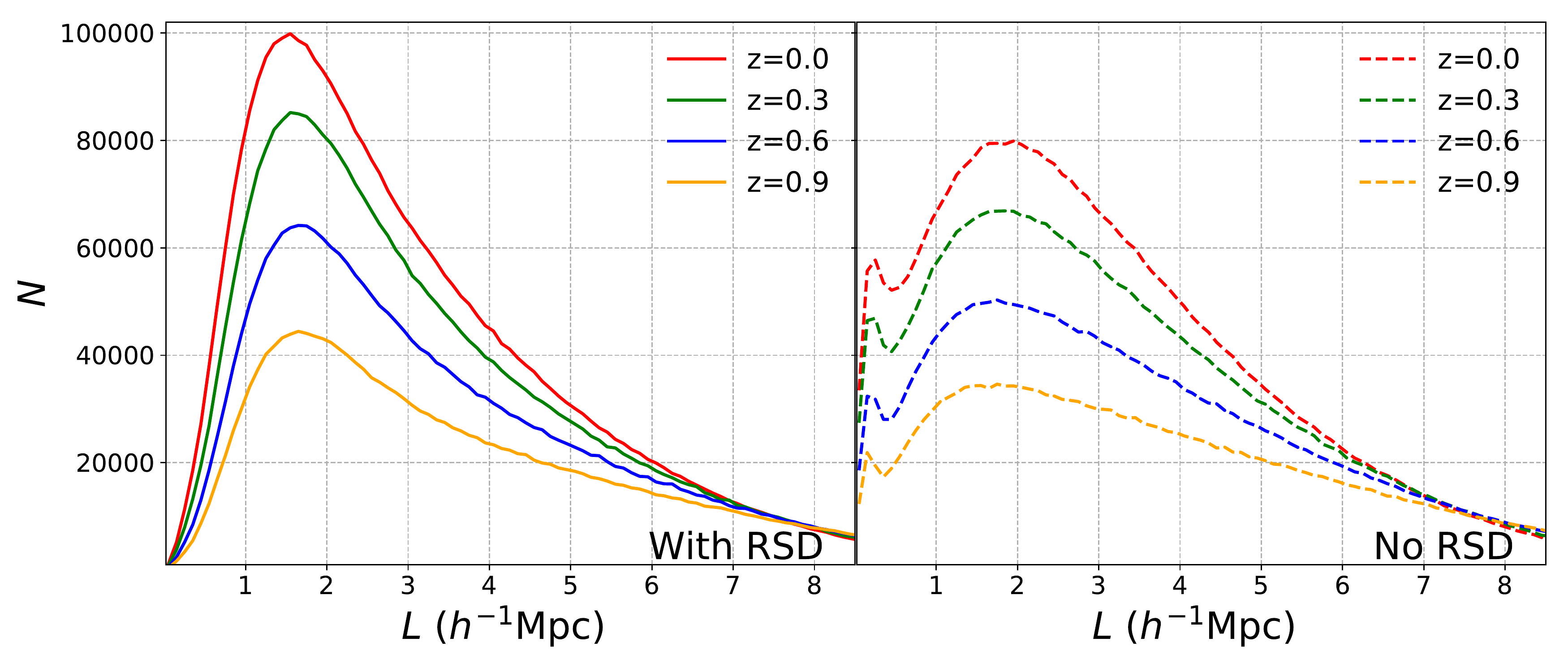}
 \includegraphics[width=10cm]{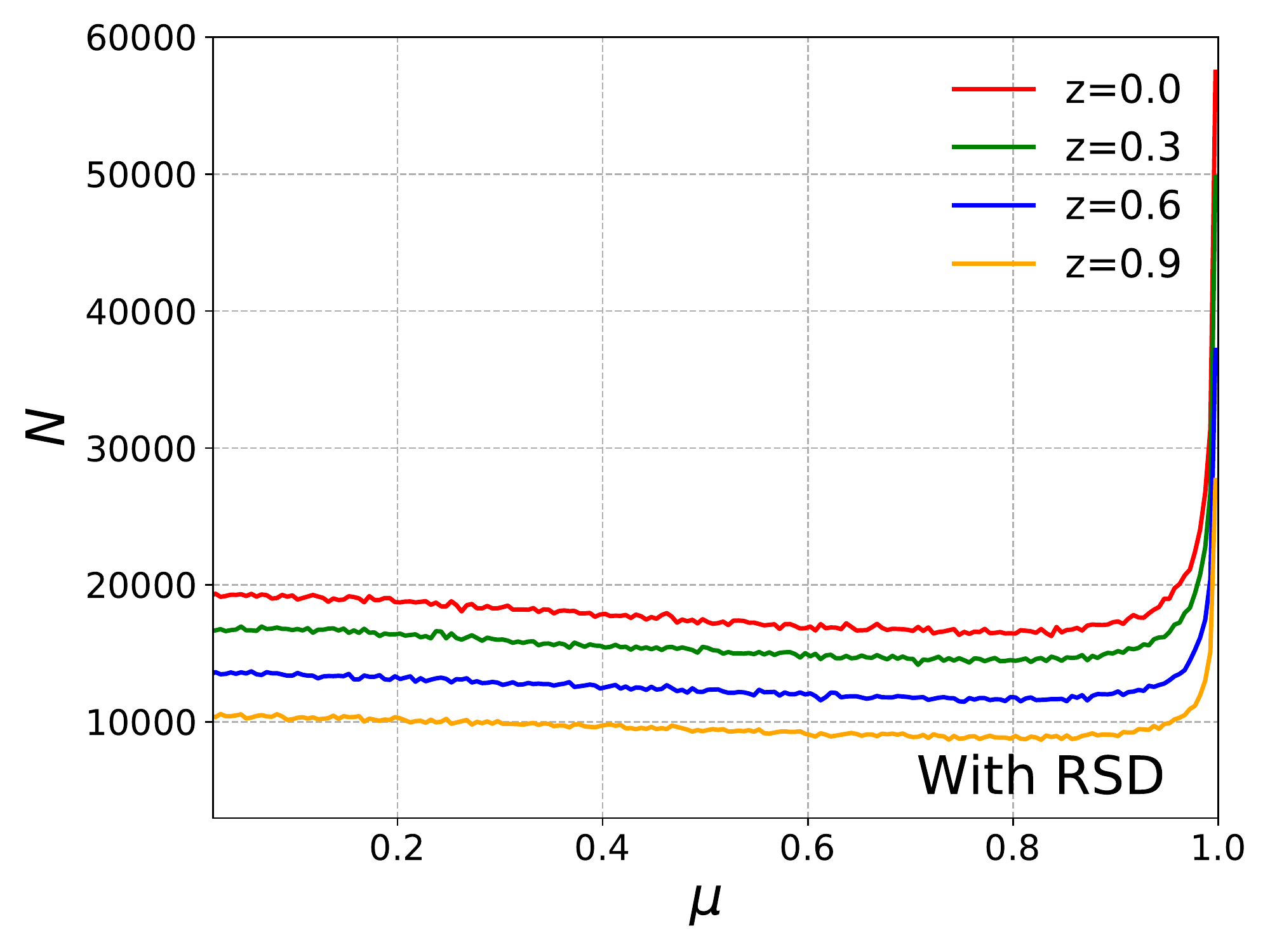}
 \caption{\label{fig_diffz}  Histograms of the connecting lengths $L$ with (top-left panel) and without (top-right panel)
 RSD effects. Those lengths are used to construct the corresponding skeleton structures, as explained in the main text. 
 The  lower panel shows the histogram of the directions of the connections $\mu$ at 4 different redshifts -- as indicated in the plot with different colors -- when
 RSD effects are included.}
\end{figure*}

Results are displayed in Figure \ref{fig_diffz}.
Specifically, the upper-right panel shows the histogram of the connecting length $L$, assuming that
the  {\it real space positions}  of galaxies are used to 
construct the skeleton (i.e., no RSD involved). 
With this assumption,  our main findings are summarized as follows:
\begin{itemize}
 \item The distribution of the connecting length peaks at $1.5-1.8 ~h^{-1} {\rm Mpc}$. This represents the typical separation length between galaxies in the skeleton. 
 Above (below) the peak scale, $N$ decreases with increasing (decreasing) $L$; a secondary peak appears at $0.1-0.2 ~ h^{-1} {\rm Mpc}$, due to the fact that there is a large number of compact clusters at this scale.
 \item As the redshift increases,
 the number of connections decreases with decreasing $N_{\rm gal}$. 
 Again, the total number of connections, which is found to be
 $3.74, 3.27, 2.62, 2.00$ at
 $z=0.9,0.6,0.3,0.0$, respectively, scales with $N_{\rm gal}$.
 \item The 4 distributions (indicated in the panel with different colors) merge at $L \approx 8 ~h^{-1}{\rm Mpc}$. 
 Above this scale, the $z=0.9$ sample shows the largest number of connections
 (even if the corresponding $N_{\rm gal}$ is significantly smaller compared to the other three samples),
 which is a clear signal that the constructed structures in this sparse sample have larger sizes -- namely, at lower redshifts, 
 objects become more compact and the distribution shifts to smaller $L$ as structures grow.
\end{itemize}

The upper-left panel in Figure \ref{fig_diffz} displays the $L$-distribution, but now using
the {\it redshift space positions} of the same galaxies considered before.
In this case,  the peculiar velocity of galaxies perturbs their observed redshifts via 
\begin{equation}\label{eq:zvpeu}
\Delta z = (1+z) \frac{v_{{\rm LOS}}}{c}, 
\end{equation}
where $v_{\rm LOS}$ is the line-of-sight (LOS) component of the velocity.
The distortion of $z$ leads to a corresponding distortion in the inferred galaxy distances, 
known as the RSD effect.
At small scales ($\lesssim 5 ~h^{-1} {\rm Mpc}$), this leads to the finger of god (FOG) feature 
\citep{jackson1972critique} (i.e., a stretch of structures along the LOS) due to 
chaotic small-scale motions of galaxies in the non-linear regime.
At large scales ($\gtrsim 40 ~ h^{-1} {\rm Mpc}$), the RSD effect is known as the `Kaiser effect' \citep{kaiser1987clustering} (i.e., a compression of structures along the LOS),
due to the coherent motions of galaxies driven by gravity.

Considering the previously reported measurements of $L$, we can infer that the  skeletons constructed from the BigMDPL  simulation 
are mainly affected by the small-scale FOG effect.
As a consequence, the number of short connections characterized by $L\lesssim 1 ~h^{-1} {\rm Mpc}$ is heavily suppressed, 
because of the stretch of distances among galaxies due to the FOG feature.
Also, 
the secondary peak -- found in the case where no RSD are considered -- now disappears.
The distribution still peaks around $1.5 ~h^{-1} {\rm Mpc}$,
but the height is $\sim 20\%$ higher than the one found in the no RSD case; this is because there is an extra contribution from the `spikes' created by the FOG effect.

Finally, the lower panel 
 in Figure \ref{fig_diffz} 
shows the histogram of $\mu$ at those four different redshifts previously specified, when RSD effects are present.
As expected, we find a non-flat distribution due to anisotropies induced by RSDs;
the FOG leads to a sharp increment of $N$ at $\mu\rightarrow 1$, and the effect is stronger at lower redshifts.

\subsection{Cosmological Effects}

We then consider the effect of cosmological parameters on the  $\beta$-skeleton statistics. 
To this end, suppose we are probing both the shape and volume of a celestial object 
by measuring its redshift span $\Delta z$ and angular size $\Delta \theta$.
We can compute its LOS dimensions in the radial ($\Delta r_{\parallel}$) and transverse ($\Delta r_{\perp}$) directions  using the relations: 
\begin{equation}\label{eq:distance}
\Delta r_{\parallel} = \frac{c}{H(z)}\Delta z,\ \ \Delta r_{\bot}=(1+z)D_A(z)\Delta \theta,
\end{equation}
where $H$ is the Hubble parameter and $D_{\rm A}$ is the angular diameter distance.
For a flat $\Lambda$CDM model with constant dark energy equation of state (DE EoS) parameter $w$,  
$H$ and $D_{\rm A}$ are simply expressed by: 
\begin{eqnarray}\label{eq:HDA}
& &H(z) = H_0\sqrt{\Omega_ma^{-3}+(1-\Omega_m)(1+z)^{3(1+w)}},\nonumber\\
& &D_A(z) = \frac{1}{1+z}r(z)=\frac{1}{1+z}\int_0^z \frac{dz^\prime}{H(z^\prime)},
\end{eqnarray}
with $H_0$ the present value of the Hubble constant, and $r(z)$ the comoving distance.

If an incorrect set of cosmological parameters is chosen in the conversion defined by Equations 
(\ref{eq:distance}) and (\ref{eq:HDA}), then the inferred $\Delta r_{\parallel}$ and $\Delta r_{\bot}$ would be both incorrect,
resulting in a distorted shape (this is known as the `AP effect') 
and in a wrongly estimated volume (this is termed as `volume effect') of the cosmological object. 
We can describe the magnitude of this combined effect via the relations: 
\begin{equation}\label{eq:stretch}
 \frac{[\Delta r_{\parallel}/\Delta r_{\perp}]_{\rm wrong}}{[\Delta r_{\parallel}/\Delta r_{\perp}]_{\rm true}} =
  \frac{[D_A(z)H(z)]_{\rm true}}{[D_A(z)H(z)]_{\rm wrong}} ,
\end{equation}
\begin{equation}\label{eq:volume}
 \frac{[\Delta r_{\parallel}(\Delta r_{\perp})^{2}]_{\rm wrong}}{[\Delta r_{\parallel}(\Delta r_{\perp})^{2}]_{\rm true}}
 = \frac{{\rm Vol}_{\rm wrong}}{{\rm Vol}_{\rm true}}
 = \frac{[D_A(z)^2/H(z)]_{\rm wrong}}{[D_A(z)^2 / H(z)]_{\rm true}},
\end{equation}
where `true' and `wrong' denote the values of those measured quantities in the actual (`true') cosmology and in the incorrectly assumed cosmology, respectively.
From the AP and the volume effects, we can therefore constrain the two quantities $D_A(z)H(z)$ and $D_A(z)^2 / H(z)$.
Clearly, these two effects will impact the statistical properties of the $\beta$-skeleton, 
which is sensitive to both the number density and the anisotropy of the cosmological sample in question.

\begin{figure*}
 \centering
 \includegraphics[width=18cm]{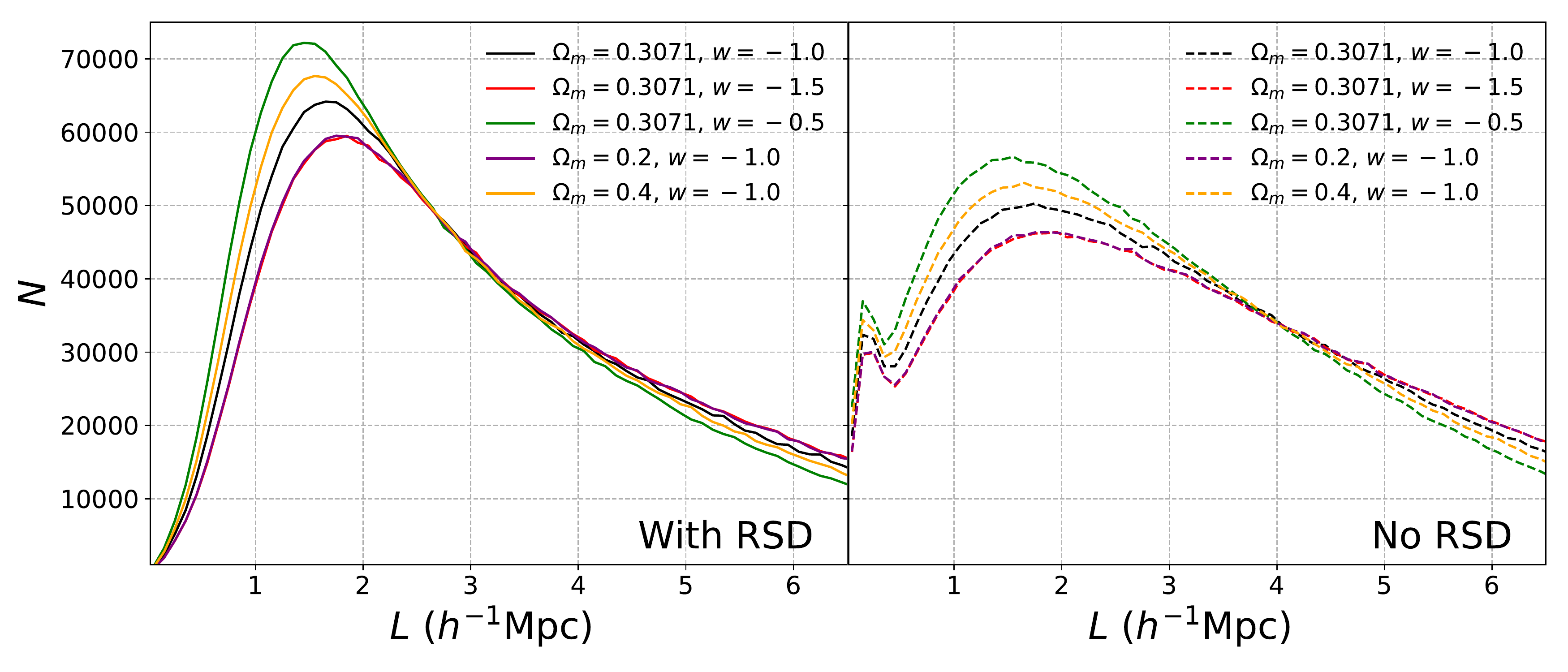}
 \caption{\label{fig_diffomw_L} {\it Sensitivity of the AP and volume effects on the $\beta$-skeleton: connecting length}. Distribution of $L$ 
 in different cosmological models defined by $\Omega_{\rm m}$ and $w$, as indicated in the various panels. The $\beta$-skeleton statistics is applied to the
 BigMDPL simulation snapshot at $z=0.6$, when RSDs are considered (redshift space -- left panel) or excluded (real space -- right panel).  
 Cosmological effects due to simultaneous variations in $\Omega_{\rm m}$ and $w$ are clearly detected.}
\end{figure*}

\begin{figure}
 \includegraphics[width=8.5cm]{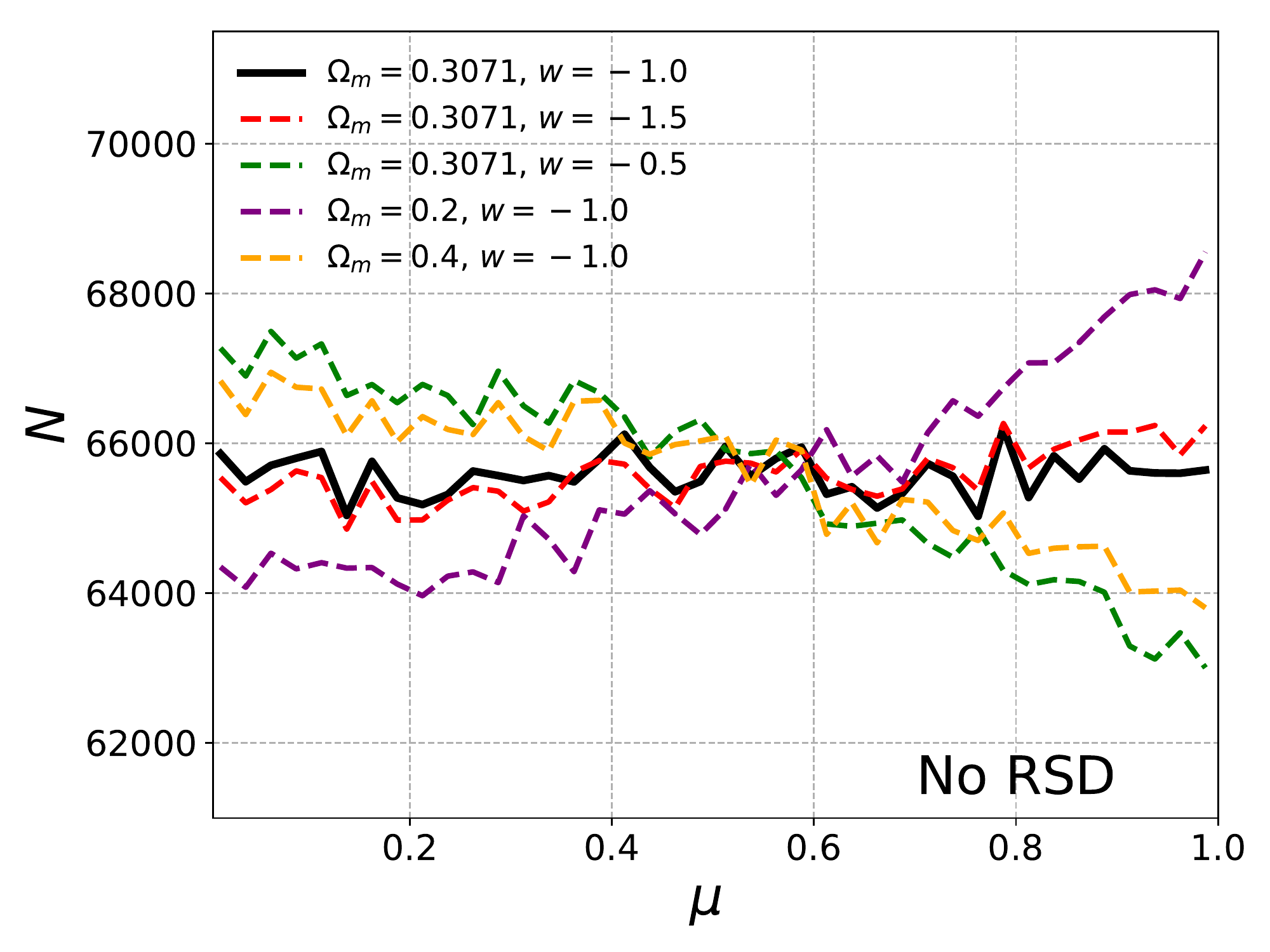}
 \caption{\label{fig_diffomw_mu} {\it Sensitivity of the AP and volume effects on the $\beta$-skeleton: connecting direction}. Distribution of $\mu$ in different cosmological models as 
in the right panel of the previous figure (no RSD effects), but now for the 
 connecting direction. See the main text for more details.}
\end{figure} 

In order to quantify the sensitivity of the AP and volume effects on the  $\beta$-skeleton,
we next apply the  $\beta$-skeleton statistics to the 
$z=0.6$ snapshot of the 
 BigMDPL sample -- but considering {\it different cosmologies}.
Namely, we 
adopt five  cosmological models characterized by 
$\Omega_{\rm m} =0.307115$ with ${\it w} = -1.0, -1.5, -0.5$, 
and $\Omega_{\rm m} = 0.2, 0.4$ with ${\it w} = -1.0$,  
and infer the actual positions of the galaxy sample using those five cosmologies in turn.  
For all those cases, we then 
analyze the statistical properties of the connection length $L$ and of the cosine of the orientation angle $\mu$. 
  
Our main results are shown in Figures \ref{fig_diffomw_L} and \ref{fig_diffomw_mu}.
Specifically,   Figure \ref{fig_diffomw_L} displays 
the histograms of $L$:   
the left panel presents the statistical distribution in redshift space (with RSDs), while the right panel shows the analogous distribution but in real space (no RSDs). 
Also in these cases, 
we find similar properties as  those highlighted in Figure \ref{fig_diffz} 
(e.g., two peaks when RSDs are not present, and one peak at $1.5~h^{-1}$Mpc if RSDs are added). 

Moreover, 
cosmological effects of varying $\Omega_{\rm m}$ and $w$ are clearly detected: 
the two cosmologies with $w=-1.5$ or $\Omega_{\rm m}=0.2$ are characterized by a faster expansion rate of the Universe compared to the `true' cosmology.
On the one hand, the comoving volume is overestimated -- resulting in a distribution of $L$ shifted to larger length; on the other hand, the overall LSS is stretched along the LOS 
because of the AP effect. Hence, 
 the distribution of $\mu$ is enhanced (suppressed) in the region where $\mu>0.5$ ($\mu<0.5$), as evident from Figure \ref{fig_diffomw_mu} -- which 
 shows only the case when RSDs are not included.\footnote{See Figure 1 of \cite{li2014cosmological,li2015cosmological} for a 
 clearer explanations on the volume and AP effects in cosmologies with incorrect $\Omega_m$ or $w$ values.}

 In the other two considered cosmologies, the effect is the opposite: 
a shrinking of the volume size shifts $L$ to smaller scales, 
and the compression of structures along the LOS tilts the distribution of $\mu$, as expected.


\subsection{Different Mass-Cut Effects}

\begin{figure*}
\centering
\includegraphics[width=9cm]{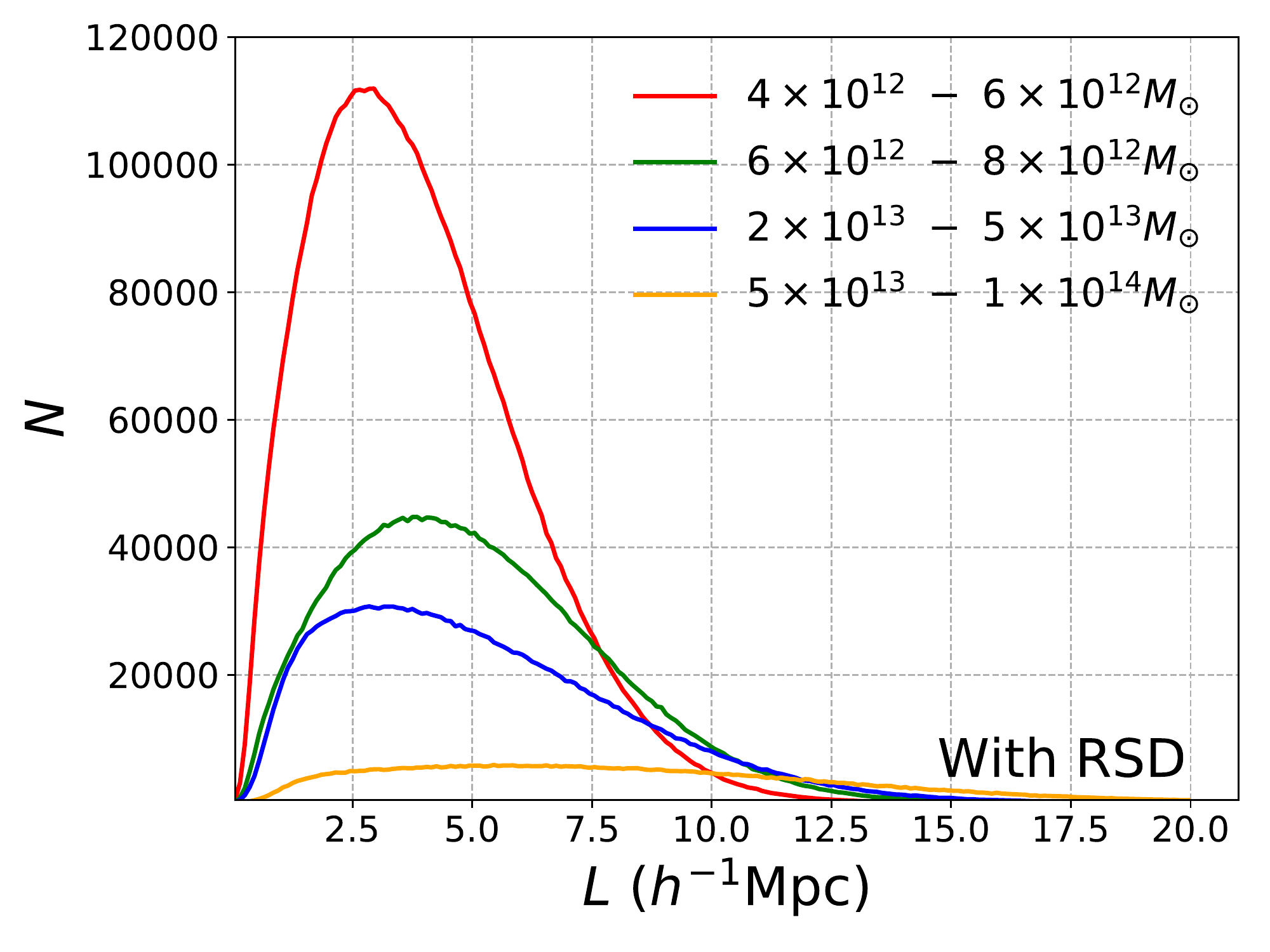}\includegraphics[width=9cm]{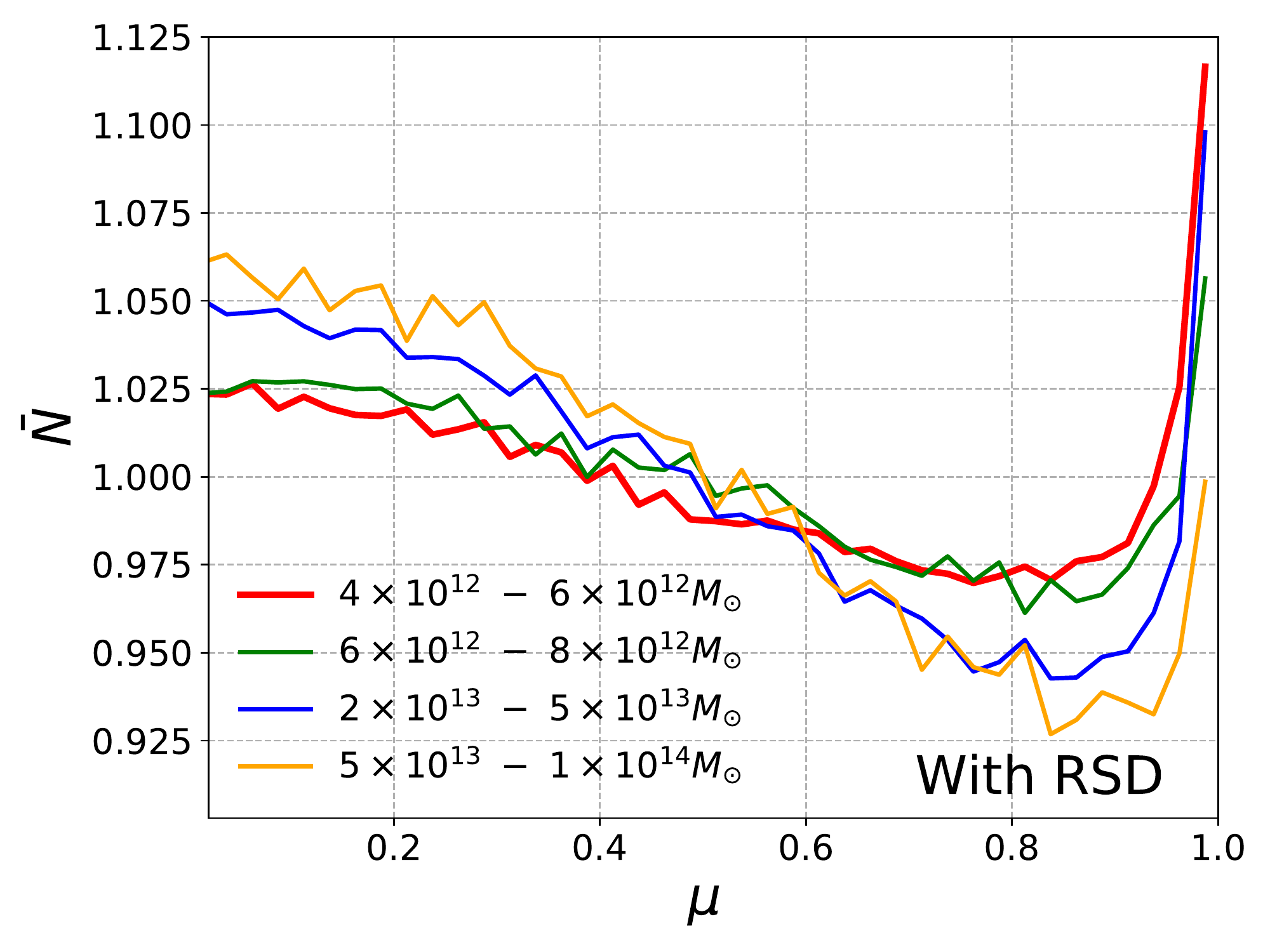}
\caption{\label{fig_diffmcut} {\it Mass-cut effects on the $\beta$-skeleton statistics}. [Left]  Distribution of the connecting length L
as a function of different mass cuts, as indicated in the panel with different line-colors, when RSDs are included and $z=0$.   [Right] Same as in the left panel, but now for the
connecting direction $\mu$.}
\end{figure*}

In the previous analysis we imposed a fixed mass-cut  to all the BigMDPL snapshots considered, namely 
$M>1\times 10^{13} M_{\odot}$. We now explore the
effect of a different mass-cut on the $\beta$-skeleton statistics. 
To this end, 
Figure \ref{fig_diffmcut} shows results of varying the mass cut (indicated with different color lines) at $z=0$, 
when RSDs are also accounted for. 
In particular, we highlight the following findings:   
\begin{itemize}
 \item When selecting galaxies in the mass intervals
[$4\times 10^{12}-6\times10^{12}] M_{\odot}$,
[$6\times 10^{12}-8\times10^{12}] M_{\odot}$,
[$2\times 10^{13}-5\times10^{13}] M_{\odot}$,
 and [$5\times 10^{13}-1\times10^{14}] M_{\odot}$,
 we obtain $N_{\rm gal}$= [6101837, 3142136, 2318205, 706430]
 and $ L= [3.96, 5.12, 5.29, 8.23] $ -- respectively.
 The relation ${\bar L} \propto N_{\rm gal}^{-1/3}$ holds well.
\item Samples characterized by a relatively smaller galaxy mass
 are dominated by satellite galaxies,
 and therefore they are more affected by the small-scale FOG effect -- resulting in a
 significant peak around $\mu \simeq 1$.
  \item On the contrary, samples with relatively larger galaxy mass 
 are more dominated by central galaxies. 
 Hence, they are more affected by the Kaiser effect, and thus present
 a more significant tilt when $\mu<0.8$.
 The peak near $\mu \sim 1$ is less significant, due to a much weaker FOG effect.
\end{itemize}


\subsection{Observational Data: Comparisons}

\begin{figure*}
 \centering
 \includegraphics[width=8.5cm]{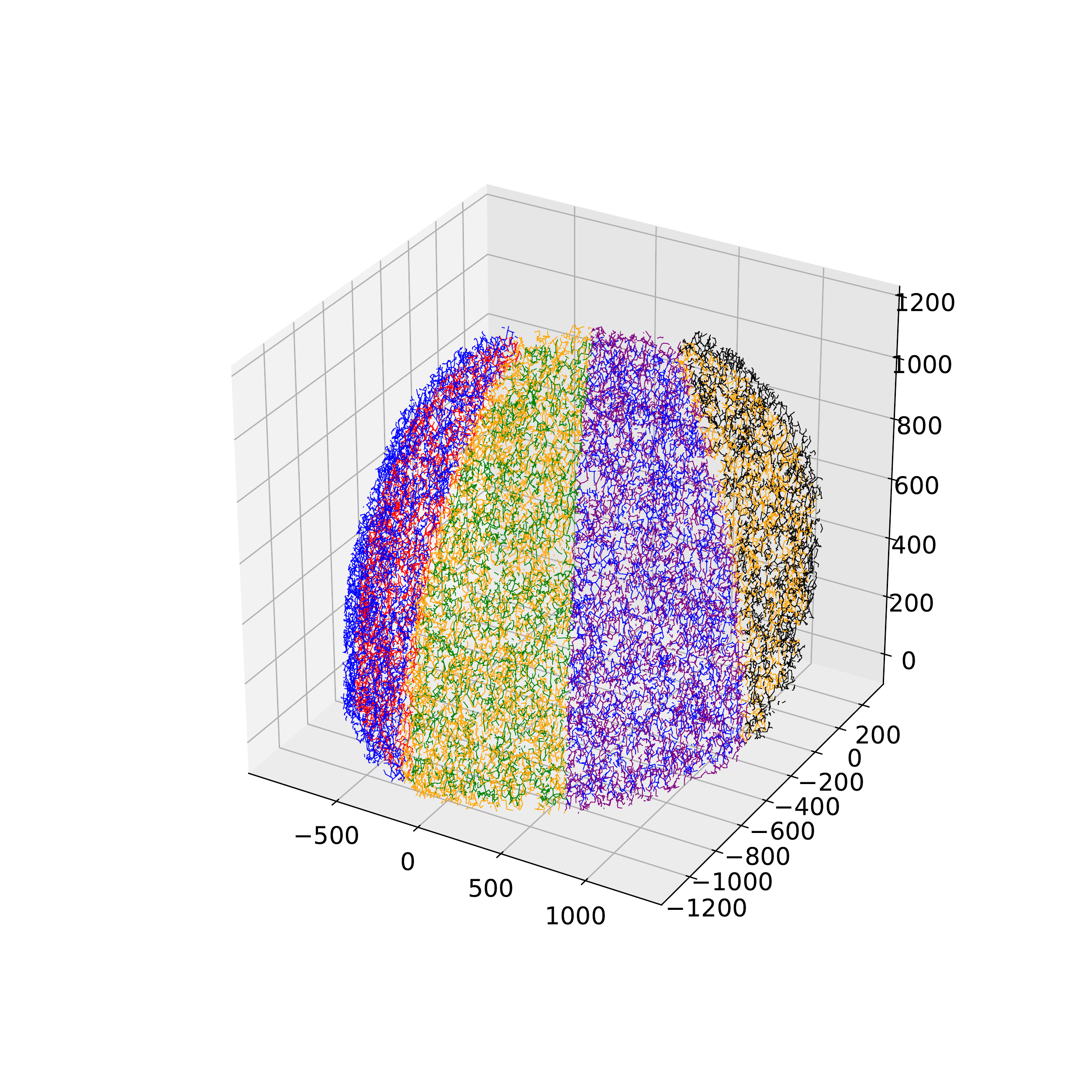}
 \includegraphics[width=7cm]{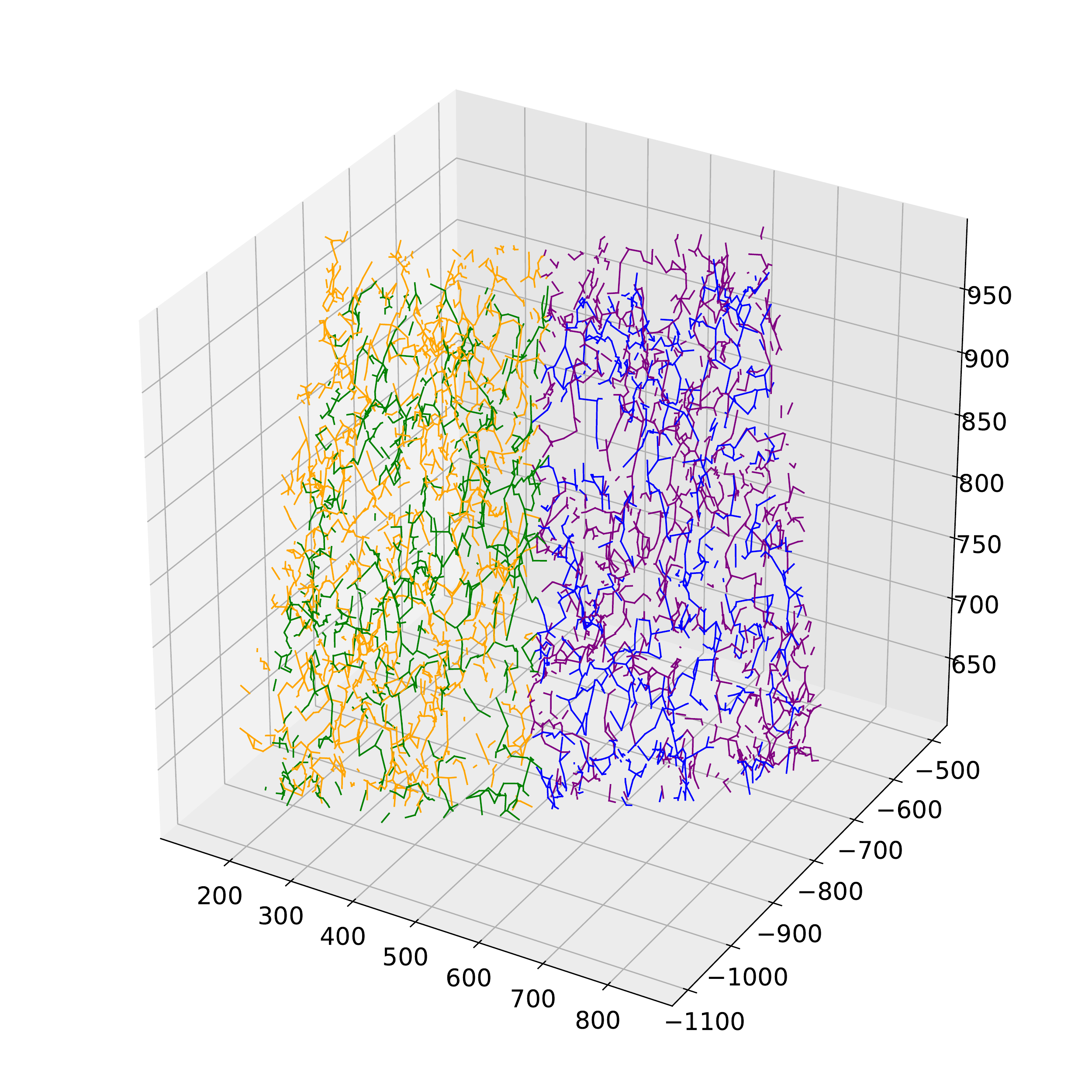}
 \caption{\label{fig_datapoints} {\it Application of the $\beta$-skeleton to observational data}. 
[Left] Visualization of the skeleton mapped by SDSS-III CMASS BOSS galaxies within the redshift range  $0.44<z<0.48$ in the  Northern sky (in units of ${h^{-1} }$ Mpc). 
Different colors indicate  galaxies with different angular directions and distances.  
[Right].   Zoom into a sub-patch of the left panel, which clearly shows the structure of the observed $\beta$-skeleton. See the main text for more details.} 
\end{figure*}

Finally, we apply the $\beta$-skeleton statistics 
to observational galaxy data, obtained from 
the SDSS-III BOSS Data Release 12 (DR12); in particular, we consider only the CMASS galaxy sample within $0.43\leq z \leq 0.7$, which contains  $\sim 0.77$ million galaxies. 
Results are shown in Figure  \ref{fig_datapoints}. Specifically, 
the left panel shows all the Northern sky galaxies in the redshift shell $0.44<z<0.48$ -- the specific redshift range has been chosen just for visualization purposes, and 
in the plot different colors indicate  galaxies with different angular directions and distances. 
The right panel is a sub-patch enlargement of the left panel, where the coordinate cut is defined by
 $170<{\rm RA}<210$ and $30<{\rm DEC}<50$.

\begin{figure*}
 \centering
 \includegraphics[width=8.5cm]{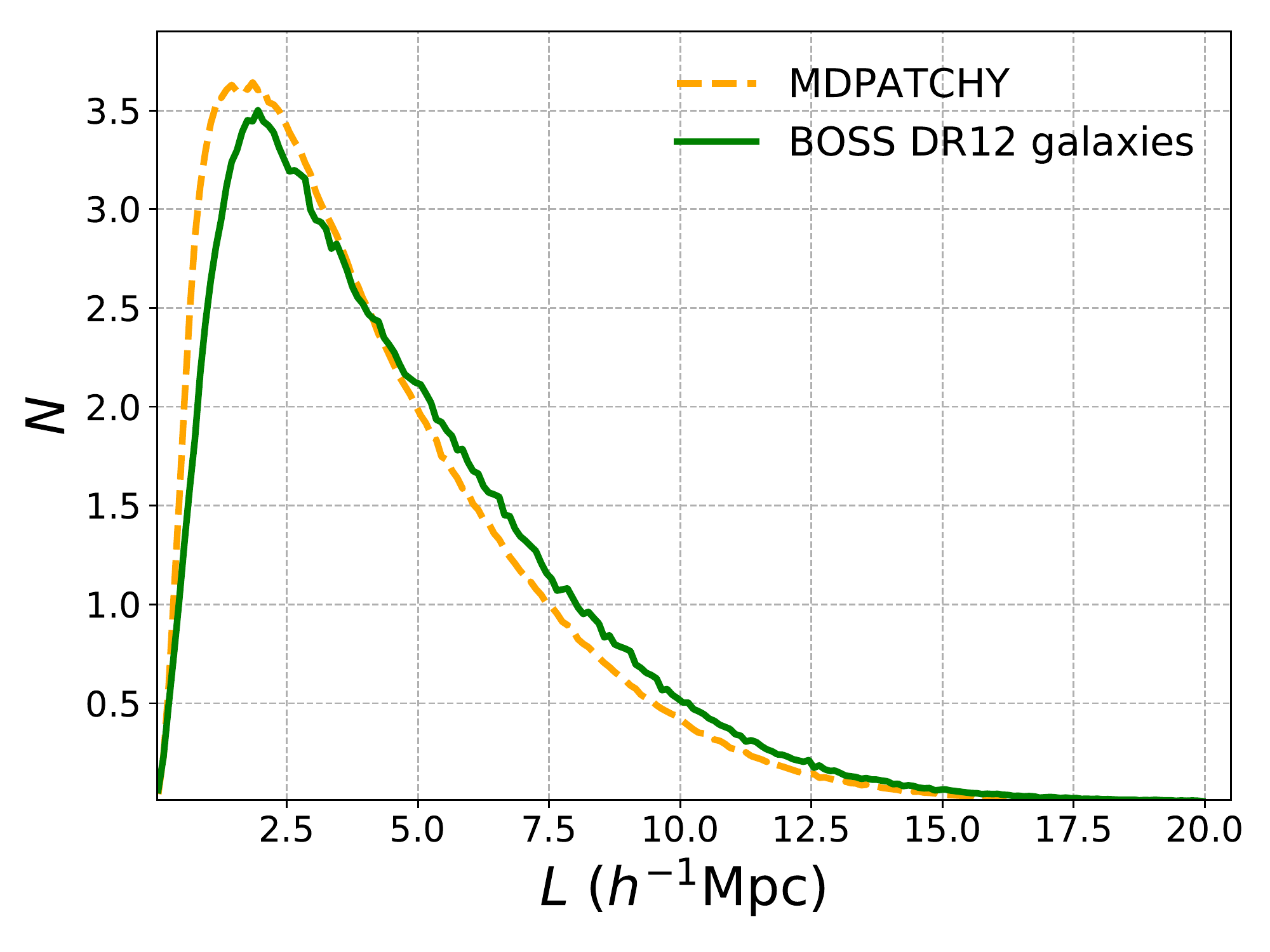}\includegraphics[width=8.5cm]{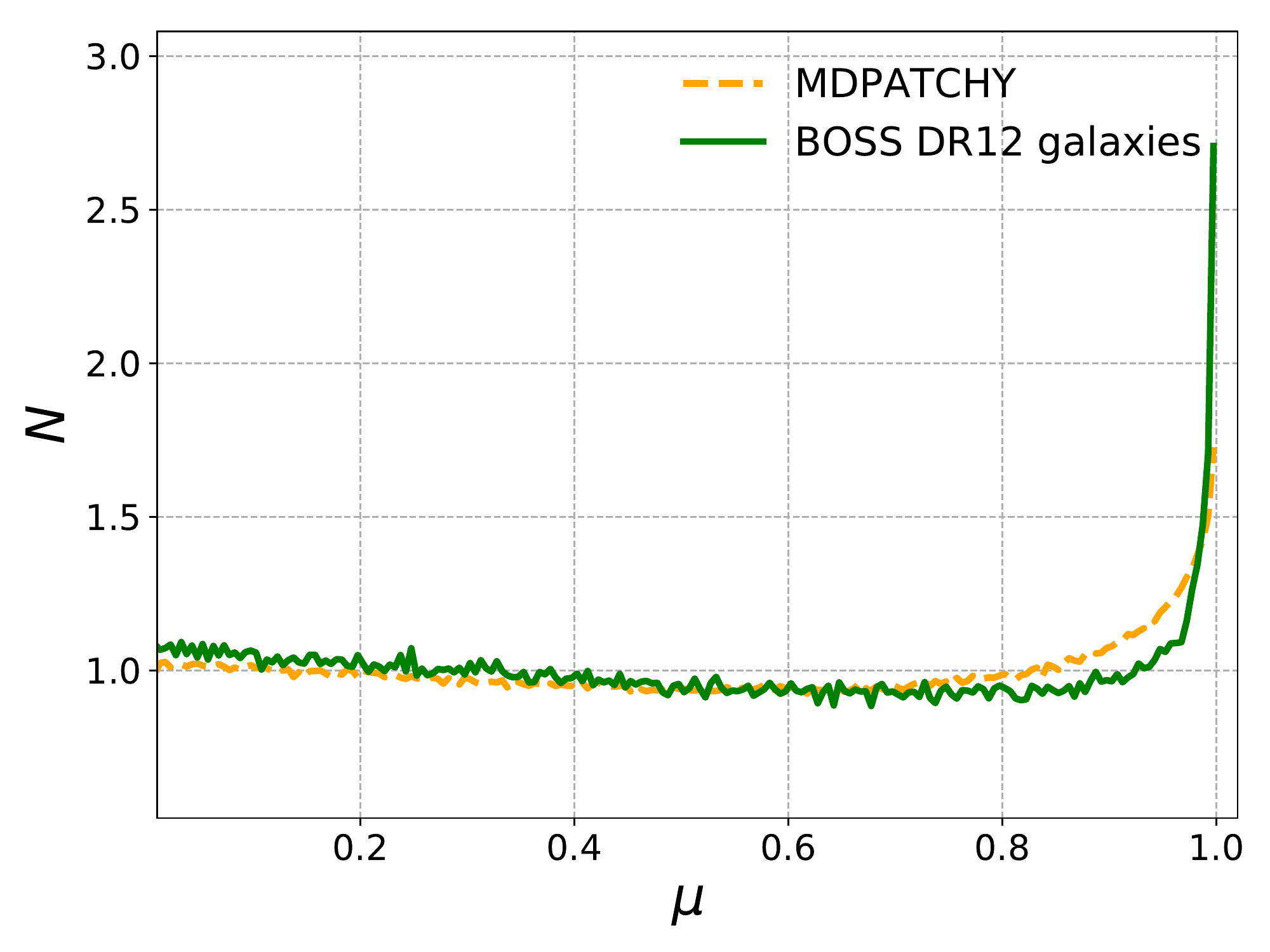}
 \includegraphics[width=8.5cm]{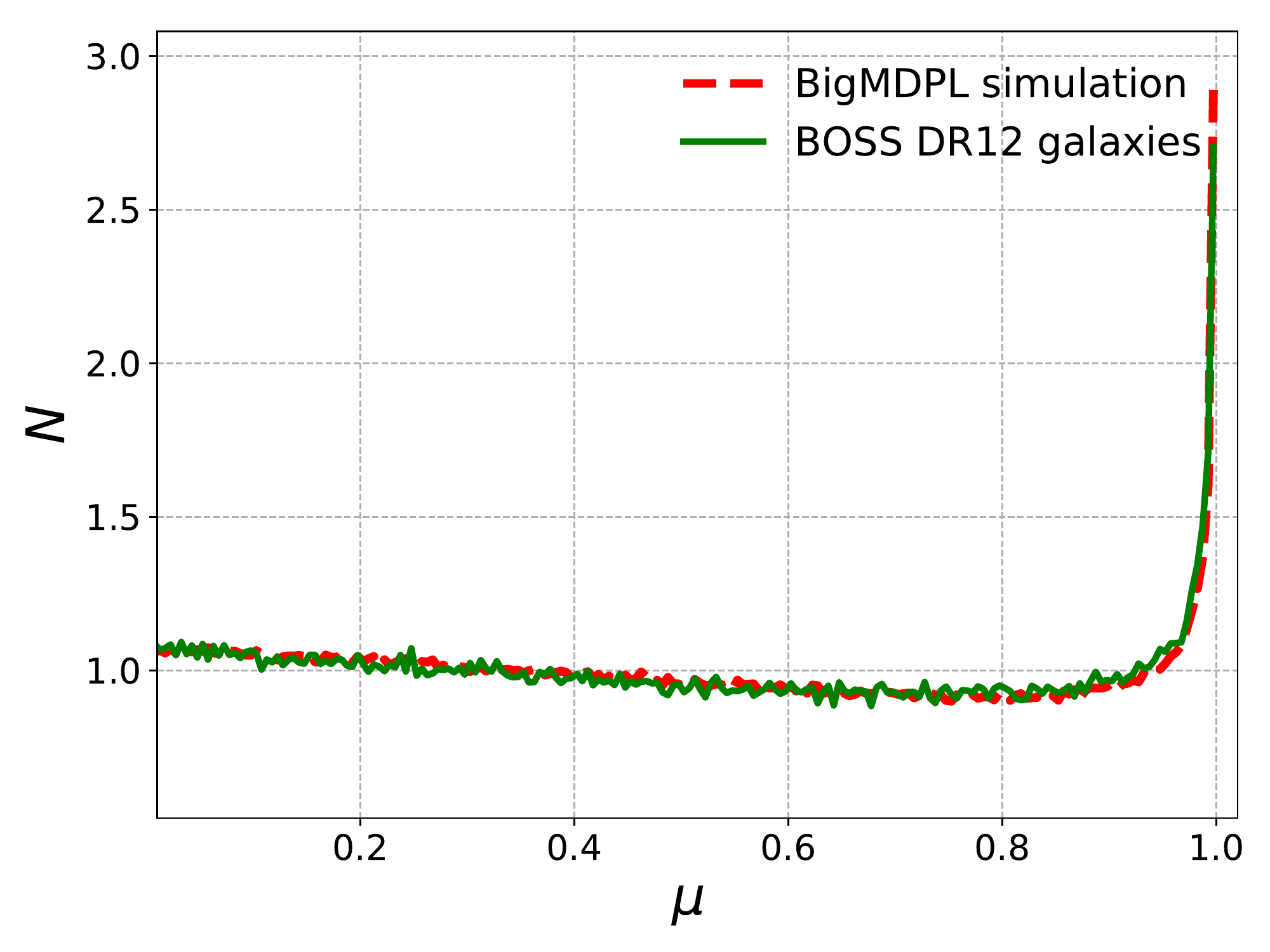}
 \caption{\label{fig_datastat} {\it Comparisons between observed and simulated $\beta$-skeleton statistics}. [Top left]   
 Connection length distributions as measured from SDSS-III BOSS galaxy data (green solid line), and as derived from the Patchy mocks (dashed yellow line). 
 [Top right] Same as in the left panel (also with identical line styles), but now for the distribution of the orientation directions; note that the Patchy mocks generally underestimate the FOG effect.   
 [Bottom]  Distribution of   $\mu$ for SDSS-III BOSS galaxies  (solid green line), and  for the BigMDPL mock at $z=0.6$ (red dashed line). The agreement between actual data and mocks 
 is better in this case, as $N$-body simulations are capable of well-reproducing the RSD effect.}
\end{figure*}

In addition, Figure \ref{fig_datastat} presents a comparison 
between the statistical properties of the $\beta$-skeleton as inferred from the SDSS-III BOSS CMASS galaxy sample,
and of  4 mock MD-PATCHY realizations that are constructed to mimic the BOSS CMASS sample, plus a 
BigMDPL snapshot at $z=0.6$. The main findings are as follows: 
\begin{itemize}
\item The observed and simulated distributions of the connection lengths $L$  are in good agreement. 
They both peak at $L \sim1.5 ~h^{-1}$ Mpc, and decrease outside of this interval. 
\item The MD-PATCHY mocks generally underestimate the FOG effect, 
a fact evident if one looks towards the $\mu\rightarrow1$ side of the lower panel in Figure \ref{fig_datastat}.
\item The $\mu$ distribution of BOSS galaxies is  much closer to the one derived from 
the BigMDPL mock at $z=0.6$, indicating that 
 $N$-body simulations are capable of well-reproducing the RSD effect present in the data.
\end{itemize}

\section{Concluding Remarks}


\subsection{Brief Summary}            

In this work, we performed a first investigation of the application of
$\beta$-skeleton statistics to cosmic web data.  
We use the BigMDPL simulation as a testing sample, and study how the
constructed skeleton depends on the values of $\beta$, redshifts, RSD,
AP and volume effects, and different and mass cuts.  
We find a significant variation of the length and direction of the
cosmic web connections under different parameters and assumptions.

We then apply the $\beta$-skeleton method to SDSS-III BOSS DR12 CMASS
galaxies, and compare our measurements with MD-PATCHY mocks.  
We find that the $N$-body sample provides a rather similar
$\mu$-distribution to the one of the data,  implying that  RSD effects
of the sample are accurately reconstructed.  
On the contrary, the MD-PATCHY mocks appear to underestimate the
magnitude of the FOG effect,  although they are designed to correctly
reproduce the $2$- and $3$-point correlation functions of the data.   

The $\beta$-skeleton clearly reveals the underlying structures encoded in the sample of points. 
From its definition, we see that  it does not require us to pre-select
a specific scale (such as the linking length in the FoF algorithm).  
One can in fact adjust the value of $\beta$, and obtain a
skeleton-like structure with different magnitudes of sparseness.  
Furthermore, the statistical properties of the $\beta$-skeleton depend
on the RSD effect, on the AP and volume effects, and on galaxy bias.  
Hence, in turn they could be used as a statistical tool to
characterize the magnitude of these effects.  


\subsection{Comparison with 2-Point Statistics}

A standard cosmological analysis generally involves the computation of the 2PCF, and of 2-point-related statistics. 
In computing the 2PCF, 
one considers all the possible pairs of galaxies (restricted to some specific scale),  and study their main clustering properties. 
Instead, the $\beta$-skeleton statistics
  focuses only on the small fraction of pairs which traces the structure; 
hence,  the physical information is actually concentrated on a subset of galaxies. 
Also, the computation of the $\beta$-skeleton is much faster than the 2PCF, so it can be used as a complementary fast statistical tool to study 
the basic properties of a given  sample. 

Although the pairs that define the $\beta$-skeleton constitute a subset of those involved in the 2PCF calculations, 
one cannot conclude that the information derived from the $\beta$-skeleton analysis is just a subset of the one inferred from 2PCF measurements. 
For example, 
Figure \ref{fig_datastat} already reveals that the MD-PATCHY mocks, constructed to reproduce the 2PCF of the data, 
have instead a rather different $\beta$-skeleton statistics from the actual data.\footnote{The $\beta$-skeleton distribution can be thought as a `weighted' 2PCF statistics, in which   
 galaxy pairs are weighted by 0 or 1, respectively, based on a graphical criterion. One may be able to extract additional information from this particular weighting scheme.}
 
This is also one main reason to pursue a $\beta$-skeleton analysis: the 2-point statistics, although powerful,
essentially compresses all the LSS information into histograms, while the cosmic web presents a much 
richer and complex  structure that can only be revealed with higher-order, more detailed analysis. 
 

\subsection{Future Investigations}

This work is a first attempt to apply the $\beta$-skeleton statistics to describe the cosmic web. 
Of course, our study can be further expanded in several directions. 
For instance, in this paper we only focused on the distribution of $L$ and $\mu$, in order to characterize  
the size and anisotropy of the LSS, but additional quantities can be used in future investigations.   
An example is represented by the 
number of connections linked at every galaxy,  which allows us to study and weight the `knots' (which connect together different filaments). 
Another possibility is to study how the connection lengths of galaxies differ depending on their environment. Namely, if they are within 
a homogeneous structure such as cluster, their connection length values should be statistically close to unity, 
while for galaxies lying at the boundary of clusters and filaments we expect those values to deviate from unity;  
the magnitude of the deviation describes how sharp the LSS are transformed from cluster-like to filament-like structures. 

Another possibility is to compare the $\beta$-skeleton statistics  with other cosmic web structure finders  -- e.g., friends-of-friends (FoF) \citep{davis1985evolution}, density-based techniques \citep{klypin1997particle,springel2001populating,knollmann2009ahf}, T-web \citep{hahn2007properties,forero2009dynamical}, V-web \citep{hoffman2012kinematic,forero2014cosmic}, etc. 
Interesting points to address include the following: finding a value of $\beta$ that yields a cosmic web realization similar to the one obtained with a different method;  
finding $\beta$ for which the connections best trace the filament-like structures identified by a different realization of the cosmic web;   
using an alternative method to classify the cosmic web into clusters, filaments, walls, and voids, and study the statistical properties of the $\beta$-skeleton in those regions;  
using the $\beta$-skeleton statistics to study how the RSD effect varies in cluster, filament, wall, and void regions; etc. 

Moreover, the $\beta$-skeleton can have several other applications in
galaxy clustering analysis -- being fast to compute and particularly sensitive to clustering properties. For example, it can be used to assess how well mocks can reproduce the properties of the observational sample,
since it is sensitive to the strength and anisotropy of clustering. It
can also be directly used  to derive quantitative constraints on
cosmological parameters, as the $\beta$-skeleton statistics are
sensitive to the AP, volume, and RSD effects. 
This could be quantified by a $\beta$-correlation that compares the
length of skeleton wedges built from data, randoms, and joint
data/randoms: that function can be defined in such a  way that in the limit
$\beta\rightarrow 0$ converges  to the two  correlation
function. 
  
Finally, in this paper we only applied $\beta$-skeleton statistics to study the LSS, but this method can be refined and developed further along with other techniques in order
to better characterize the properties of the cosmic web, and extract useful cosmological information.  


\section*{Acknowledgements}

J.E. F-R acknowledges support from COLCIENCIAS Contract No. 287-2016,
Project 1204-712-50459.  
 G.R. acknowledges support from the National Research Foundation of
 Korea (NRF) through Grant No. 2017R1E1A1A01077508 funded by the
 Korean Ministry of Education, Science and Technology (MoEST), and
 from the faculty research fund of Sejong University in 2018. 
F.L.L. acknowledges support from  Key Program of National Natural Science Foundation of China (NFSC) through grant 11733010 and 11333008,  and the State Key Development Program for Basic Research of China (2015CB857000).
  
 We greatly acknowledge Changbom Park for many helpful discussions.
 

\bibliographystyle{mnras}

\end{document}